\documentclass[aps,12pt]{revtex4}
\usepackage{epsfig}
\newcommand{\be}{\begin{equation}}
\newcommand{\ee}{\end{equation}}
\newcommand{\ba}{\begin{array}}
\newcommand{\ea}{\end{array}}

\begin{document}

\title{Phase Transition Properties of 3D Potts Models}
\author{Alexei Bazavov$^{\,a}$, Bernd A. Berg$^{\,b,c}$ and
        Santosh Dubey$^{\,b,c}$ }
\affiliation{$^{a)}$ Department of Physics, University of Arizona,
                  Tucson, AZ 85721, USA \\
$^{b)}$ Department of Physics, Florida State University, 
             Tallahassee, FL 32306, USA \\
$^{c)}$ School of Computational Science, Florida State University, 
             Tallahassee, FL 32306, USA }
%\date{\today}
\date{April 8, 2008}

\begin{abstract} 
Using multicanonical Metropolis simulations we estimate phase 
transition properties of 3D Potts models for $q=4$ to~10: The
transition temperatures, latent heats, entropy gaps, normalized 
entropies at the disordered and ordered endpoints, interfacial
tensions, and spinodal endpoints.

\end{abstract}

\maketitle
\baselineskip = 18pt

\section{Introduction} \label{sec_Intro}

Potts models were introduced as a footnote in the thesis by Potts 
\cite{Po52} on clock models, whose investigation had been proposed 
to him by his adviser Domb. We define their energy function by
\begin{equation} \label{PottsE}
  E^{(k)} = 2\, \sum_{\langle ij\rangle} \left(\frac{1}{q} -
  \delta(q_i^{(k)},q_j^{(k)})\right)
\end{equation}
where $\delta(q_i,q_j)$ is the Kronecker delta function, the sum 
$\langle ij\rangle$ is over the nearest neighbor lattice sites, the 
index $(k)$ refers to the configuration, and $q_i^{(k)}$ is the Potts 
spin at site $i$. For the $q$-state model Potts spins take the values 
$0,1,\dots,q-1$. Following the notation of Ref.~\cite{BBook} the factor 
of two in front of the sum is introduced to match for $q=2$ on Ising 
model conventions. Expectation values of the Gibbs canonical ensemble 
are calculated with the Boltzmann (or Gibbs) factor $\exp(-\beta E)$, 
$\beta=1/(kT)$.

Potts, and not so much clock models, received considerable attention 
up to the day. Developments till 1982 were reviewed by Wu~\cite{Wu82}. 
In 2D the models have second order phase transitions for $q=2$ to 4, 
and first order transitions for $q\ge 5$. In a work by Baxter 
\cite{Ba73}, critical 
temperatures, latent heats and entropies were analytically derived. 
Equations for interface tensions followed later~\cite{BoJa92} and
played a role in the verification of the multicanonical approach
\cite{BeNe92} to Markov Chain Monte Carlo (MCMC) simulations. In 3D 
the $q=2$ Ising model phase transition is second and the $q\ge 3$ 
transitions are first order. The strength of the first order 
transitions, measured by their latent heat, increases with $q$.
The $q=5$ transition in 2D and the $q=3$ transition in 3D are
weakly first order.

Potts models proved amazingly versatile to grasp the essence of 
physically interesting situations, many of them are described in 
Ref.~\cite{Wu82}. When they are generalized by introducing quenched 
random variables as exchange constants, the Ising case of the 
Edwards-Anderson spin glass \cite{EdAn75} and quadrupolar Potts 
glasses~\cite{Bi97} are obtained. Particular choices give
anti-ferromagnets and the fully frustrated Ising model~\cite{Vi77}.

The 3D $q=3$ Potts model shares the center symmetry of SU(3) gauge
theory \cite{SY82} and plays a role in our understanding of the QCD 
deconfining phase transition. This and other motivations led to a 
number of numerical investigations of the 3D 3-state Potts model 
\cite{GKB89,FMOU90,ABV91,Schm94,JV97,KaSt00,Fa07,BaBe07}.
However, when we recently looked out for a strong first order 
transition in 3D, we found only a few papers \cite{GN02,Ha05,HJ06} 
dealing with 3D Potts for $q\ge 4$. The purpose of 
this paper is to fill this gap in the literature for 3D Potts models 
up to $q=10$. Thereby, we will closely follow the outline of a previous 
investigation of the equilibrium statistical properties of the 3D 
3-state Potts model by two of the present authors~\cite{BaBe07}.

Next, we briefly summarize our simulation method and give an overview 
of our assembled statistics. In section~\ref{sec_Ana1} we calculate
and analyze transition temperatures, specific and latent heats. 
Section~\ref{sec_Ana2} deals with entropy and energy across the
phase transition, section~\ref{sec_Ana3} with interface tensions
and spinodal endpoints. A brief summary and conclusions are
given in the final section~\ref{sec_sum}.

\section{Simulation Method and Overview of Data} \label{sec_Data}

We want to calculate expectation values $\mathcal{O}$ in the Gibbs 
canonical ensemble. They are given by the ensemble average
\begin{equation} \label{O}
  \widehat{\mathcal{O}} = \widehat{\mathcal{O}} (\beta) = 
  \langle \mathcal{O} \rangle_{\beta}  = Z^{-1} \sum_{k=1}^K
  \mathcal{O}^{(k)}\,e^{-\beta\,E^{(k)} }
\end{equation}
where
\begin{equation} \label{Z}
 Z = Z(\beta) = \sum_{k=1}^K e^{-\beta\,E^{(k)} } 
   = \sum_E n(E)\,e^{-\beta\,E} 
\end{equation}
is the partition function. The index $(k),\, k=1,\dots , K$ labels 
the configurations (microstates) of the system and $E^{(k)}$ is 
the (internal) energy of configuration $(k)$. In the last equation 
$n(E)$ is the number of configurations with energy $E$.

We consider Potts models on cubic lattices of size $L^3$ with periodic
boundary conditions. There are $N=L^3$ Potts spins. Each microstate
$k$ defines a unique arrangement of Potts spins and vice versa:
\begin{equation} \label{k}
 k = \{ q_1^{(k)}, \dots , q_N^{(k)} \} \ . 
\end{equation}
As each Potts spin can take on $q$ values, there are
\begin{equation} \label{K}
  K_L = Z_L(0) = \sum_E n_L(E) = q^N\ ,  
\end{equation}
microstates. 
Even for rather small numbers of $L$, $K$ is a very large, so that 
one will not be able to sum the partition function explicitly. 
Instead, one can use statistical methods. 

MCMC simulations \cite{BBook} are a suitable approach to evaluate
equilibrium properties of the canonical ensemble. Off the phase 
transition temperatures canonical Metropolis or heatbath simulations 
with weight factor $\exp(-\beta E)$ work sufficiently well, provided 
a ordered start is used for simulations in the ordered phase, and 
(somewhat less important) a disordered start for simulations in the 
disordered phase. However, canonical simulations deteriorate quickly 
when it comes to the (most interesting) investigation of phase 
transition properties. For first order transitions the reason is 
that the relevant transition states are in the canonical ensemble 
suppressed $\sim \exp(-2\sigma_{od}\,L^{D-1})$, where $\sigma_{od}$ 
is the order-disorder interface tension.

Multicanonical simulations \cite{BeNe92,BBook} are a remedy for this 
supercritical slowing down. One performs for $E_{\min}\le E\le E_{\max}$ 
MCMC simulations with a working approximation of the weight factors
\begin{equation} \label{MUCAweights}
  w_{\rm muca}(E) = e^{-b(E)\,E+a(E)} = \frac{1}{n(E)}\,,
\end{equation}
supplemented by the canonical weights $\exp(-\beta_{\max} E)$ for 
$E<E_{\min}$ and $\exp(-\beta_{\min} E)$ for $E>E_{\max}$. Here $n(E)$ 
is the number of states with energy $E$ as introduced by Eq.~(\ref{Z}). 
With the weights (\ref{MUCAweights}) all energies in the range 
$E_{\min}\le E\le E_{\max}$ are sampled with the same probability, 
so that the Markov process will perform some kind of random walk in 
the range $E_{\min}\le E\le E_{\max}$. The microcanonical inverse
temperature $b(E)$ and the dimensionless free energy $a(E)$ follow
uniquely from $n(E)$ due to the relation
\begin{equation} \label{micro_beta}
  a(E-\epsilon) = a(E) +\left[b(E-\epsilon)-b(E)\right]\,E\,,
  ~~~a(E_{\max})=0\,,
\end{equation}
where $\epsilon$ is the step to the next energy. However, $n(E)$ is 
a-priori unknown, so that some iterative procedure needs to be used. 
In this context working approximation of the weights means that any 
weights that enable cycling (also called tunneling)
\begin{equation} \label{cycle}
  E_{\max}\ \to\ E_{\min}~~~{\rm and~~back}
\end{equation}
are considered to be acceptable. Actually it is known that not the 
weights (\ref{MUCAweights}), but some modifications of them, can be
optimal for that purpose~\cite{HS95,THT04} and it should be noted that 
there is a residual exponential slowing down~\cite{NH03}.

We find working estimates of the multicanonical weights by finite size 
(FS) extrapolations from a smaller lattice to a next larger lattice, 
a method which was already used in~\cite{BeNe92}. Obviously, this 
requires that the FS behavior of the system under consideration is 
well defined. In particular, for complex systems like spin glasses 
or proteins this is not the case and more sophisticated recursion 
approaches need to be used \cite{BBook}, most noted is presumably 
the one by Wang and Landau \cite{WL01}. 

Let $e=E/N=E/L^3$ be the energy density. For our Potts models we simply 
convert the microcanonical inverse temperature $b_{L_1}(E)$ for a given 
lattice size $L_1$ to the $b_{L_2}(E)$ of our next larger lattice size 
$L_2>L_1$ via the interpolation
\begin{equation} \label{rho}
  b_{L_2}(E) = w_+\,b_{L_1}(E_+) + w_-\,b_{L_1}(E_-)~~~{\rm with}~~~
  w_{\pm} = \frac{|e-e_{\mp}|}{e_+-e_-}\ , 
\end{equation}
where $e=E/(L_2)^3$ and $e_-<e$ and $e_+>e$ are the values closest 
to $e$, so that we have for the corresponding $E_+$ and $E_-$ values
entries in $b_{L_1}(E)$. 
For our purposes this simple procedure turned out to be sufficiently 
accurate. Better extrapolations can be expected by taking into 
account details of the shape of $n_L(E)$, as for a magnetic field
driven phase transition discussed in Ref.~\cite{BHN93}.

\begin{table}
\centering
\caption {\label{tab_stat}{Analyzed statistics per production run
in sweeps: $32\ \times$ the number given.}} \medskip
\begin{tabular}{|c|c|c|c|c|c|c|c|}
\hline\hline
$L$&   $q=4$       &  $q=5$ & 		$q=6$ & 	 $q=7$ & 	  $q=8$ & 	   $q=9$ & 	    $q=10$ \\
\hline
$\beta_{\max}=$&0.45&0.41&0.43&0.45&0.47&0.49&0.50\\ \hline
  2&  256& 		1024&   	 6$\times 10^3$ & 10$\times 10^3$& 15$\times 10^3$& 15$\times 10^3$& 16384 \\
\hline
  3&  1024& 		2048&       	 9$\times 10^3$ & 10$\times 10^3$& 15$\times 10^3$& 15$\times 10^3$&  4000 \\
\hline
  4&  1024& 		2048&       	 9$\times 10^3$ & 10$\times 10^3$& 20$\times 10^3$& 20$\times 10^3$&  4000\\
\hline
  6&  4096& 		8192&       	 1$\times 10^4$ & 15$\times 10^3$& 20$\times 10^3$& 30$\times 10^3$&  65536 \\
\hline
  8&  4096& 		32768&       	 4$\times 10^4$ & 45$\times 10^3$& 5$\times 10^4$ & 75$\times 10^3$&  262144 \\
\hline
 10&  16384& 		10$\times 10^4$& 16$\times 10^4$& 25$\times 10^4$& 3$\times 10^5$ & 4$\times 10^5$ &  1048576\\
\hline
 12&  65536& 		25$\times 10^4$& 3$\times 10^5$  & 45$\times 10^4$& 55$\times 10^4$& 7$\times 10^5$ &  4194304\\
\hline
 14&  262144&		45$\times 10^4$& 5$\times 10^5$ & 14$\times 10^5$& 17$\times 10^5$& 22$\times 10^5$&  16777216\\
\hline
 16&  262144& 		85$\times 10^4$& 15$\times 10^5$& 20$\times 10^5$& 26$\times 10^5$& 32$\times 10^5$&  2$\times 10^7$\\
\hline
 18&  1048576&		13$\times 10^5$& 21$\times 10^5$& 29$\times 10^5$& 70$\times 10^5$& 2$\times 10^7$ &  4$\times 10^7$\\
\hline
 20&  4194304&		19$\times 10^5$& 30$\times 10^5$& 50$\times 10^5$& 14$\times 10^6$& 8$\times 10^7$ &  8$\times 10^7$ \\
\hline
 22&  $-$    & 		$-$            & $-$            & 1$\times 10^7$ & 3$\times 10^7$ & $-$            &  $-$           \\
\hline 
 24&  45$\times 10^5$&	39$\times 10^5$& 1$\times 10^7$ & 3$\times 10^7$ & $-$            & $-$		   &  $-$      \\
\hline
 26& 9$\times 10^6$&  	$-$&       	 2$\times 10^7$ & $-$            & $-$            & $-$            &  $-$    \\
\hline
 28&14$\times 10^6$&	1$\times 10^7$&  $-$            & $-$            & $-$            & $-$            &  $-$    \\
\hline
 30&15$\times 10^6$&    3$\times 10^7$&  $-$            & $-$            & $-$            & $-$            &  $-$      \\
\hline \hline
\end{tabular}
\end{table}

Suitably, the $E_{\max}$, $E_{\min}$, $\beta_{\min}$ and $\beta_{\max}$
parameters, which accompany the weights (\ref{MUCAweights}) are chosen
so that they embrace the phase transitions and that
\begin{equation} \label{EmaxEmin}
  E_{\max} = \langle E\rangle_{\beta_{\min}}\,, \qquad
  E_{\min} = \langle E\rangle_{\beta_{\max}}
\end{equation}
holds. From a simulation with these weights, canonical expectation
values are obtained by reweighting for the temperature range 
$\beta_{\min}\le\beta\le\beta_{\max}$. This property has coined 
the name multicanonical. To calculate the partition function (\ref{Z}), 
from which the normalized entropy and free energy follow, one has 
to include $\beta=0$ in this temperature range. So we choose 
$\beta_{\min}=0$, for which our normalization of the energy 
(\ref{PottsE}) implies $E_{\max}=0$.
For $\beta_{\max}$ we chose the values given in table~\ref{tab_stat},
each of them well above the transition value $\beta_c$. Our simulations 
of multicanonical ensembles defined by the weights (\ref{MUCAweights}) 
rely on the Metropolis algorithm. We update sequentially with one 
Metropolis update per spin during one sweep through the lattice. 
This is more efficient \cite{BBook} than picking spins at random 
for the updates. 

\begin{table}
\centering
\caption {\label{tab_tun1}{Number of cycling events for the first 
production runs. }} \medskip
\begin{tabular}{|c|c|c|c|c|c|c|c|}
\hline\hline
      & $q=4$ & $q=5$ & $q=6$ & $q=7$ & $q=8$ & $q=9$ & $q=10$ \\
\hline
$L=~2:$&    3 &  6       &  4       &  2     & 1      & 1   & 1     \\
\hline
$L=~3:$&   32 &  135     &  350     & 246    & 239    & 104 & 15     \\
\hline
$L=~4:$&  109 &  139     &  377     & 358    & 550    & 531 & 88     \\
\hline
$L=~6:$&  68  &  117     &  80      & 98     & 89     & 87  & 164     \\
\hline
$L=~8:$&  21  &  125     &  106     & 76     & 71     & 92  & 113     \\
\hline
$L=10:$&  20  &  186     &  190     & 189    & 122    & 109 & 135     \\
\hline
$L=12:$&  77  &  249     &  195     & 169    & 99     & 52  & 107     \\
\hline
$L=14:$&  173 &  279     &  178     & 242    & 102    & 42  & 86     \\
\hline
$L=16:$&  125 &  313     &  301     & 139    & 47     & 12  & 11     \\
\hline
$L=18:$&  353 &  322     &  228     & 69     & 38     & 10  & 3     \\
\hline
$L=20:$&  711 &  327     &  172     & 48     & 11     & 5   & 2     \\
\hline
$L=22:$&  $-$ &  $-$     &   $-$    & 27     & 6      & $-$ & $-$     \\
\hline
$L=24:$&  441 &  84      &   1      & 27     & $-$    & $-$ & $-$     \\
\hline
$L=26:$&  903 &  $-$     &  116     & $-$    & $-$    & $-$ & $-$     \\
\hline
$L=28:$&  867 &  43      &  $-$     & $-$    & $-$    & $-$ & $-$     \\
\hline
$L=30:$&  356 &  639     &  $-$     & $-$    & $-$    & $-$ & $-$     \\
\hline \hline
\end{tabular}
\end{table}

In table~\ref{tab_stat} we give an overview of the statistics per run. 
We followed the outline of Potts model MCMC simulations in~\cite{BBook}. 
First, we performed the number of sweeps listed in table~\ref{tab_stat} 
for reaching equilibrium. Data from these sweeps are excluded from the 
statistics for which measurements were performed. Subsequently, we 
collected for each run 32 histograms, each relying on the number of 
sweeps listed in the table. All error bars are then calculated with 
respect to these 32 bins (32 jackknife bins when nonlinear operations 
on the data are involved). From the student distribution it is known 
that error bars from 32 independent Gaussian data give almost Gaussian 
confidence probabilities at the level of two standard deviations.

For most data points we performed two runs with the statistics of
table~\ref{tab_stat}. The first runs are based on the weights iterated
from the closest smaller lattice. These data are taken to refine the 
weights for the lattices at hand. The refined weights are used for 
the second production runs on these lattices. Exceptions from this 
procedure are iterations from a smaller to a larger lattice
immediately after the first run. This speeds up the process of
getting to larger lattices and has often been done when the 
cycling frequency of the first run was already satisfactory.

\begin{table}
\centering
\caption {\label{tab_tun2}{Number of cycling events for the second
production runs. }} \medskip
\begin{tabular}{|c|c|c|c|c|c|c|c|}
\hline\hline
      & $q=4$ & $q=5$ & $q=6$ & $q=7$ & $q=8$ & $q=9$ & $q=10$ \\
\hline
$L=~2:$&  139 & 1451  & 5918   & 7091   & 2614     & 7985     & 6339      \\
\hline
$L=~3:$&  288 & 378   & 1275   & 1167   & 1436     & 1209     & 1098     \\
\hline
$L=~4:$&  72  & 126   & 410    & 331    & 578      & 497      & 324     \\
\hline
$L=~6:$&  69  & 100   & 109    & 127    & 117      & 165      & 311     \\
\hline
$L=~8:$&  22  & 154   & 125    & 117    & 105      & 110      & 349     \\
\hline
$L=10:$&  42  & 180   & 223    & 295    & 243      & 273      & 598     \\
\hline
$L=12:$&  93  & 256   & 221    & 228    & 218      & 246      & 1142     \\
\hline
$L=14:$&  209 & 251   & 219    & 427    & 384      & 394      & 2423     \\
\hline
$L=16:$&  130 & 308   & 366    & 396    & 352      & 359      & 1542     \\
\hline
$L=18:$&  323 & 301   & 347    & 342    & 585      & 1056     & 1707     \\
\hline
$L=20:$&  921 & 307   & 329    & 358    & 590      & 2168     & 1591     \\
\hline
$L=22:$&  $-$ & $-$   & $-$    & 405    & 779      & $-$      & $-$     \\
\hline
$L=24:$&  521 & 321   & 487    & 910    & $-$      & $-$      & $-$     \\
\hline
$L=26:$&  $-$ & $-$   & 708    & $-$    & $-$      & $-$      & $-$     \\
\hline
$L=28:$&  986 & 460   & $-$    & $-$    & $-$      & $-$      & $-$     \\
\hline
$L=30:$&  869 & $-$   & $-$    & $-$    & $-$      & $-$      & $-$     \\
\hline \hline
\end{tabular}
\end{table}

Table~\ref{tab_tun1} collects the number of cycling (\ref{cycle}) 
events obtained in the first production runs. For $L=2$, some cycling 
is already achieved by a canonical simulation at $\beta=0$. This 
allows one to determine multicanonical weights for the second run 
on $2^3$ lattices, which have large cycling rates as shown in 
table~\ref{tab_tun2}, and to start off FS iterations of the weights. 
From the $2^3$ lattices we extrapolate weights for the first runs on 
$3^3$ lattices, refine them for the second runs on $3^3$ lattices, 
iterate to the next larger lattice, and so on (up to the before 
mentioned exceptions). The calculations were carried out on PC 
clusters at FSU. Our present lattice sizes are limited by the 
deterioration of cycling with increasing $L$, the computational 
power of a single PC (here 2-3 GHz per PC), and the limitation 
of the total length of one run to a few months.

During the simulations we collect histograms of the energy in the 
multicanonical ensemble, $h_{mu}(E)$, and calculate functions of the
energy $f(E)$ from them by reweighting to the canonical ensemble:
\begin{equation} \label{f_muca}
  \overline{f} = { \sum_E f(E)\, h_{mu}(E)\, \exp 
  \left[ -\beta\,E + b(E)\,E-a(E) \right] \over \sum_E h_{mu}(E)\, 
  \exp \left[ -\beta\,E + b(E)\,E-a(E) \right] }\ ,
\end{equation}
where the sums are over all energy values for which $h_{mu}(E)$ has 
entries. The computer implementation of this equation requires care,
because the differences between the largest and the smallest numbers 
encountered in the exponents can be large. We rely here on the 
logarithmic coding of Ref.~\cite{BBook}. Whenever the function
$f(E)$ is non-linear jackknife binning is employed.

\section{Energies, Transition Temperatures and Latent Heats} 
\label{sec_Ana1}

\begin{figure} \begin{center}
\epsfig{file=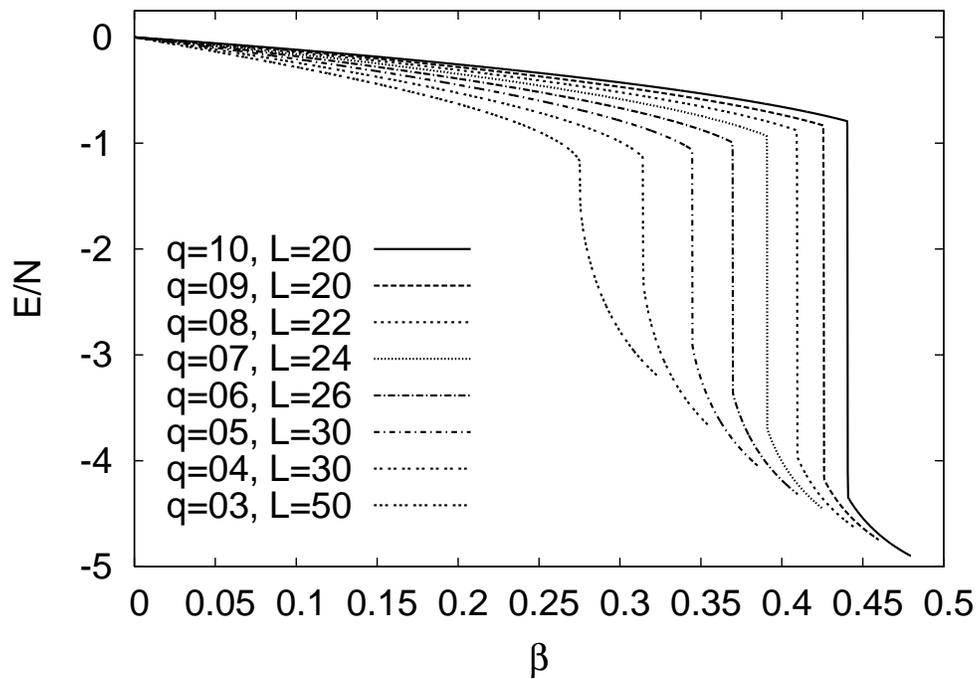,width=0.82\textwidth, angle=0}
\caption{Internal energies. } \label{fig_energy} 
\end{center} \end{figure}

In Fig.~\ref{fig_energy} we show internal energies as functions of 
$\beta$, for each $q$ from the largest available lattice (error bars 
are not resolved on the scale of this figure). This gives a rough 
estimate of the latent heats and the inverse transition temperatures 
$\beta_t$. Accurate results follow from FS extrapolations of 
indicators, which are defined on finite lattices, so that they 
converge in the limit $L\to\infty$ (quickly) towards the infinite 
volume value of the desired physical quantity.

We calculate specific heats via the fluctuation-dissipation theorem
\begin{equation} \label{C}
  C = \frac{(\beta)^2}{L^3}\left(\langle E^2\rangle -
      \langle E\rangle^2\right)\ .
\end{equation}
For first order phase transitions the finite volume specific heats 
are regularization of Dirac delta functions, which are the infinite 
volume extrapolations. The multicanonical approach allows to 
calculate specific heat values for a continuous range of $\beta$ 
values. Consequently, the locations of the maxima can be accurately 
determined.

Finite lattice indicators for the transition temperatures are called 
pseudo transition temperatures, and there are various options to 
define them. On finite volumes their values differ, while they all 
converge to the same $L\to\infty$ limit. We use here three definitions 
of pseudo transition temperatures: $\beta^1_{pt}$, the $\beta$ value 
at which equal heights are achieved in the double peaked energy 
histogram, $\beta^2_{pt}$, the position of the central energy of 
the latent heat, and $\beta^3_{pt}$, the location of the maximum of 
the specific heat. The first two definitions are explained in more 
detail later, $\beta^2_{pt}$ plays a role in determining the entropy 
gaps in section~\ref{sec_Ana2} and $\beta^1_{pt}$ for the interface 
tensions in section~\ref{sec_Ana3} (they are labeled in this order 
for consistency with Ref.~\cite{BaBe07}). 

\begin{figure} \begin{center}
\epsfig{file=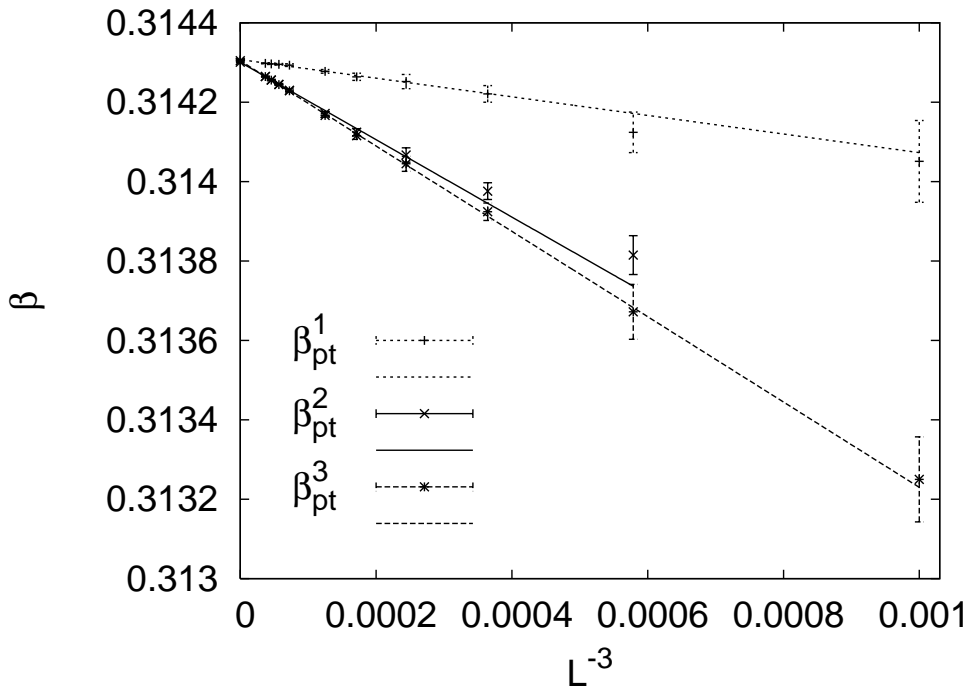,width=0.82\textwidth, angle=0}
\caption{Transition temperature fits for $q=4$.} \label{fig_beta04q} 
\end{center} \end{figure}

For $q=4$ Fig.~\ref{fig_beta04q} shows our linear fits 
\begin{equation} \label{beta_fits}
  \beta^i_{pt}(L) = \beta^i_t + \frac{c^i}{L^3}\ ,
\end{equation}
which determine the infinite volume transition temperatures. The 
vertical order of the fits agrees with that in the legend. While the 
finite volume estimators $\beta^i_{pt}(L)$ differ, the infinite volume
extrapolations are consistent with one another. The smallest lattices
have been omitted from the fits to ensure an acceptable goodness 
of fit $Q$~\cite{BBook} in each case. This $q=4$ pattern repeats for 
all $q$: The quality of the fits from our three definitions of pseudo 
transitions temperatures are similar, and the final estimates as well 
as their error bars are consistent with one another. To give one best 
number for each $q$, we simply average over the three estimates. We 
average also their error bars, because all three estimators rely on 
the same simulation, so that one does not expect error bar reduction 
when averaging over them. The thus obtained transition values 
$\beta_t(q)$ are collected in table~\ref{tab_estimates1}. For the 
convenience of the reader we have included $q=3$ and~2 estimates 
from Ref.~\cite{BaBe07} and \cite{TB96}, respectively. So far we 
have not found a simple formula for the $q$ dependence like 
$\beta_t=\ln(1+\sqrt{q})/2$, which holds in 2D \cite{Ba73} (given 
here in our convention). 

\begin{table}
\centering
\caption{\label{tab_estimates1}{Estimates of observables ($q=2$ 
from Ref.~\cite{TB96} and $q=3$ from Ref.~\cite{BaBe07}).}} \medskip
\begin{tabular}{|c|c|c|c|c|c|}
\hline\hline
 $q$ & $\beta_t$    & $\triangle e$ & $e(\beta_t)$  & $e^+$ & $e^-$ 
\\ \hline
 $2$ & 0.2216544 (06) &  0           & -0.9957 (14) & $e(\beta_t)$  &
       $e(\beta_t)$ \\ \hline
 $3$ & 0.2752827 (29) &  0.3286 (15) & -1.3470 (74) &  -1.1826 (73) & 
      -1.5112 (79) \\ \hline
 $4$ & 0.3143041 (17) & 1.16294 (61) & -1.719 (36)  & -1.1367 (64) & 
      -2.3019 (64) \\ \hline
 $5$ & 0.3447205 (12) & 1.84619 (20) & -1.987 (20)  & -1.063 (20)&
       -2.910 (20) \\ \hline
 $6$ & 0.3697070 (15) & 2.36442 (17) & -2.177 (26)  & -0.995 (26) &
     -3.359 (26) \\ \hline
 $7$ & 0.3909657 (17) & 2.76430 (12) & -2.316 (22)  & -0.934 (22) &
      -3.698 (22) \\ \hline
 $8$ & 0.4094959 (23) & 3.08039 (15) & -2.421 (32)  & -0.881 (32) &
      -3.961 (32) \\ \hline
 $9$ & 0.4259432 (23) & 3.33628 (12) & -2.503 (32)  & -0.835 (32) &
     -4.171(32) \\ \hline
$10$ & 0.4407371 (18) & 3.547570 (87)& -2.567 (25)  & -0.794 (25) &
      -4.341 (25) \\ \hline
\hline 
\end{tabular}
\end{table}

Following Ref.~\cite{CLB86}, we extract the latent heat $\triangle e 
= \triangle E/N$ by fitting the maxima of the specific heat to the form
\begin{equation} \label{C_fit}
  C_{\max}(L) = a_1 + a_2\,L^3
\end{equation}
and using the relation
\begin{equation} \label{latent_heat}
  a_2 = \frac{1}{2}\,(\beta_t)^2\,(\triangle e)^2\ .
\end{equation}
The results are also included in table~\ref{tab_estimates1}.

\section{Entropy and Energy Across the Phase Transitions} \label{sec_Ana2}

The entropy density is
\begin{equation}\label{entr_den}
  s=\beta\,(e-f)\ ,
\end{equation}
where $f$ is the free energy density, which is continuous at the phase 
transition. So, the entropy gaps across the phase transitions are
\begin{equation}\label{Sgap}
  \triangle s=\beta_t\,\Delta e\,,
\end{equation}
or $\triangle s=2\sqrt{a_2}$ with $a_2$ from the fits (\ref{C_fit}).
The entropy and energy density endpoints in the disordered ($+$) and 
ordered ($-$) phases are given by
\begin{eqnarray}\label{splus_def}
  s^+ &=& \beta_t\,(e^+-f(\beta_t))\,,\\
  s^- &=& \beta_t\,(e^--f(\beta_t))\,,
\end{eqnarray}
and are more difficult to compute than the gaps, because the additive 
normalization constants no longer drop out. We follow the method of 
Ref.~\cite{BaBe07}, which relies on the definition of
$\beta^2_{pt}$ as given below.

\begin{figure} \begin{center}
\epsfig{file=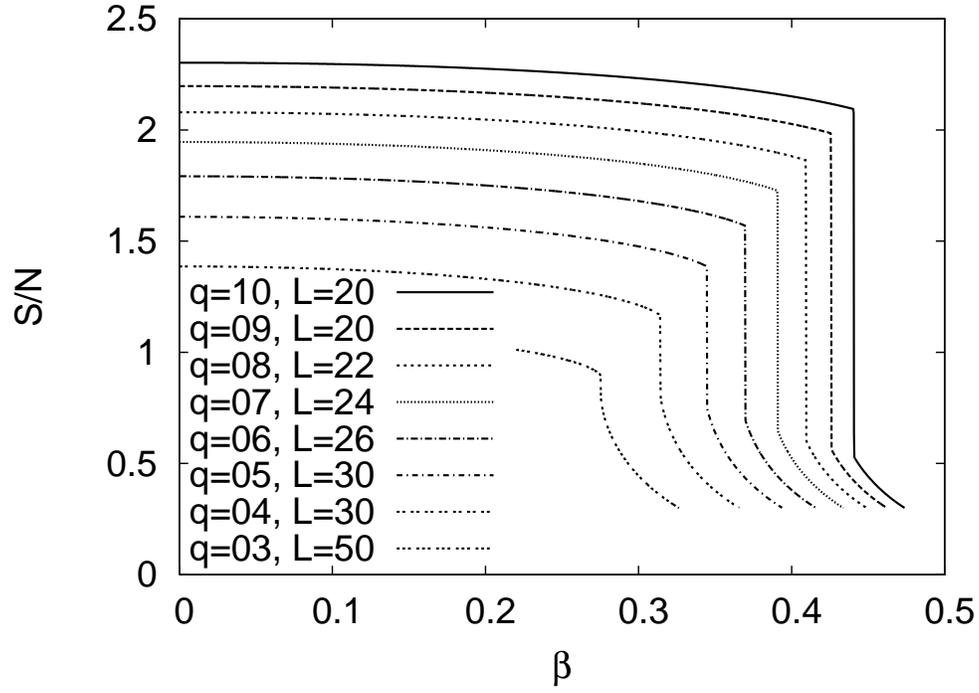,width=0.82\textwidth, angle=0}
\caption{Normalized entropy densities.} \label{fig_entropy} 
\end{center} \end{figure}

In multicanonical simulations, the normalization constant for the 
entropy is determined by the known value at $\beta=0$:
\begin{equation} \label{S0}
  S_0=\ln\left(q^N \right)~~{\rm and}~~
  s_0=\frac{S_0}{N}=\ln q\ .
\end{equation}
Fig.~\ref{fig_entropy} shows normalized entropy densities of our models
for our largest lattices. 

\begin{figure} \begin{center}
\epsfig{file=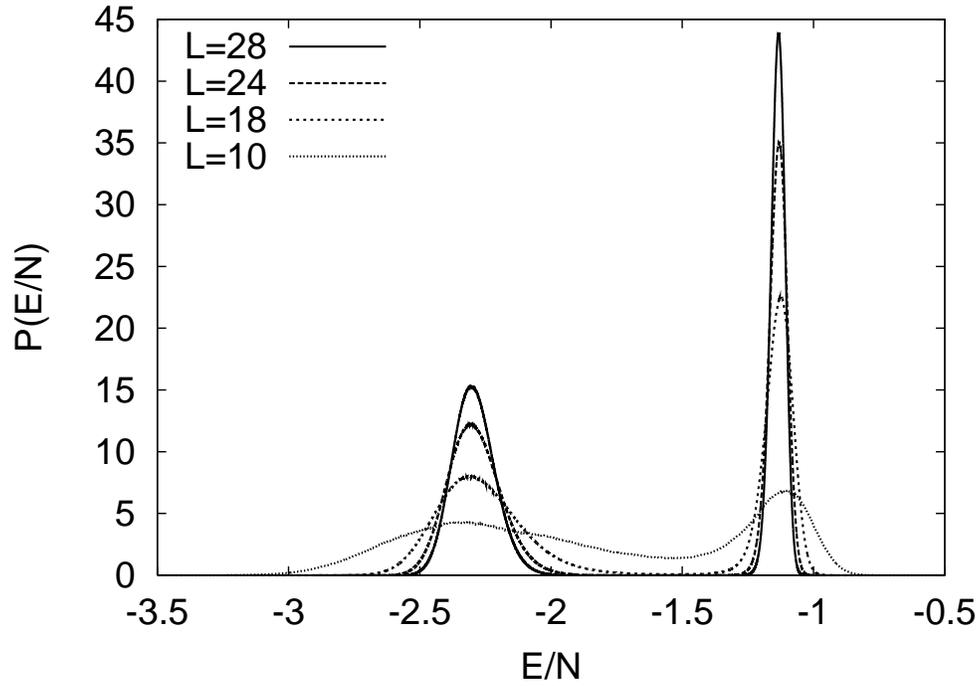,width=0.82\textwidth, angle=0}
\caption{Energy density histograms at $\beta^2_{pt}(L)$ for $q=4$.} 
\label{fig_hbeta2} 
\end{center} \end{figure}

\begin{table}
\centering
\caption{\label{tab_estimates2}{Estimates of observables 
($q=3$ from Ref.~\cite{BaBe07}).}} \medskip
\begin{tabular}{|c|c|c|c|c|c|c|}
\hline\hline
 $q$ &$s(\beta_t)$ & $f(\beta_t)$  & $s^+$ & $s^-$ & $w^+$ & $w^-$ 
\\ \hline
 $2$ & 0.55715 (31)& -3.50956 (23) & $s(\beta_t)$ &$s(\beta_t)$&
     1.75 & 1.75 \\ \hline
 $3$ & 0.8491 (21) & -4.431364 (50)& 0.8943 (21) & 0.8038 (22) &
     2.45 & 2.24 \\ \hline
 $4$ &0.983 (11)   & -4.846358 (41)& 1.166 (11)  & 0.800 (11) &
     3.21 & 2.23 \\ \hline
 $5$ & 1.0680 (70) & -5.084679 (34)& 1.3862 (70) & 0.7498 (70) &
     4.00 & 2.12 \\ \hline
 $6$ & 1.1324 (93) & -5.239933 (50)& 1.5695 (93) & 0.6954 (93) &
     4.80 & 2.00 \\ \hline
 $7$ & 1.1858 (85) & -5.349346 (60)& 1.7262 (85) & 0.6455 (85) &
     5.62 & 1.91 \\ \hline
 $8$ & 1.233 (13) & -5.430753 (66) & 1.863 (13) & 0.602 (13) &
     6.44 & 1.83 \\ \hline 
 $9$ & 1.274 (14) & -5.493609 (42) & 1.984 (14) & 0.563 (14) &
     7.27 & 1.76 \\ \hline
$10$ &  1.312 (11) & -5.543856 (50)& 2.094 (11) & 0.530 (11) &
      8.11 & 1.70 \\ \hline
\hline 
\end{tabular}
\end{table}

To calculate the endpoints of the entropy and energy on the ordered
and disordered sides of the transitions, we define $\beta^2_{pt}(L)$ 
by the relation
\begin{equation} \label{eL}
  e_L(\beta^2_{pt}) = \frac{1}{2}\,\left[ e_L^+(\beta^2_{pt})
                    + e^-_L(\beta^2_{pt}) \right]\,,
\end{equation}
where $e^{\pm}_L(\beta^2_{pt})$ are the locations of the maxima of
the double peak histogram at $\beta^2_{pt}(L)$. For $q=4$ these
histograms are shown in Fig.~\ref{fig_hbeta2} (we excluded the 
$L=30$ lattice to keep a reasonable scale in the figure). This 
construction ensures that the energy endpoints $e^{\pm}_L$ are 
positioned symmetrically about the central energy density 
$e_L(\beta^2_{pt})$:
\begin{equation} \label{epm}
  e^{\pm}_L = e_L(\beta^2_{pt}) \pm \frac{1}{2}\triangle e_L
\end{equation}
and one finds that
\begin{equation} \label{spm}
  s^{\pm}_L = s_L(\beta^2_{pt}) \pm \frac{1}{2}\triangle s_L
\end{equation}
holds as well. As in \cite{BaBe07} we use jackknife estimators and 
arrive at the values for $e(\beta_t)$, $e^+$ and $e^-$ compiled in 
table~\ref{tab_estimates1} and those for $s(\beta_t)$, $f(\beta_t)$, 
$s^+$ and $s^-$ compiled in table~\ref{tab_estimates2}. Although 
there are simple relations between these values, we have to list 
them separately, because these relations do not determine error bars. 
The estimates are correlated and the jackknife procedure takes care 
of correct error bars. Also it should be noted that $\beta_t$ in 
the arguments of $e$, $s$ and $f$ is $\beta^2_t$, defined as the 
extrapolation of the pseudo transition temperatures as defined by 
(\ref{eL}). These values are consistent with the $\beta_t$ values 
listed in table~\ref{tab_estimates1}, but not identical, as $\beta_t$ 
of table~\ref{tab_estimates1} is the average of the extrapolations 
from our three definitions of pseudo transition temperatures.

In the last two columns of table~\ref{tab_estimates2} we give
the first few digits of the effective number of states on the
disordered and ordered sides of the phase transition,
\begin{equation} \label{wpm}
  w^{\pm} = \exp(s^{\pm})\ .
\end{equation}
Amazingly, the effective number of states per spin at the the
ordered endpoint goes down by increasing~$q$.

\section{Interface Tensions and Spinodal Endpoints} \label{sec_Ana3}

\begin{figure} \begin{center}
\epsfig{file=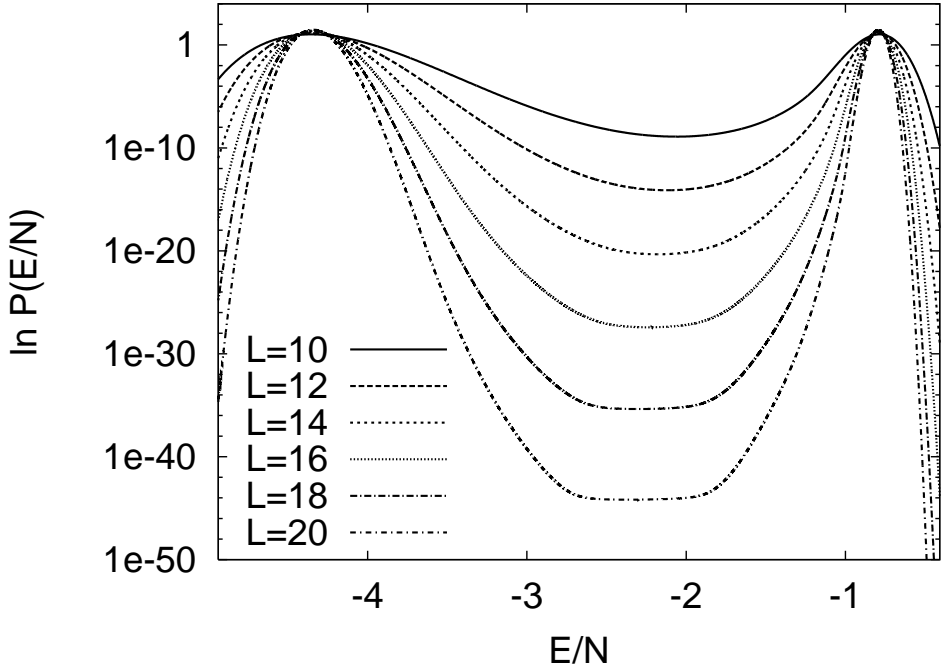,width=0.82\textwidth, angle=0}
\caption{Equal heights energy density histograms $q=10$.} 
\label{fig_heqh} 
\end{center} \end{figure}

For $L\to\infty$ the interface tension between ordered and disordered
phases is~\cite{Bi82}
\begin{equation}\label{sigma}
  2 \sigma_{od}(L) = \frac{1}{L^2}\ln
  \left(\frac{P_{\max}(L)}{P_{\min}(L)}\right)
\end{equation}
where $P_{\max}(L)$ represents the value of the maxima when the energy
histogram is reweighted to equal heights and $P_{\min}(L)$ the minimum
in between the peaks. For $q=10$ we show our equal heights histograms
in Fig.~\ref{fig_heqh}. Including capillary waves \cite{BZ,GF,Mo}, we 
perform 2- and 3-parameter fits to the form (compare Eq.~(16) of 
\cite{BNB94})
\begin{equation}\label{sigma_fit}
  2\,\sigma_{od}(L) + \frac{\ln(L)}{2L^2} =
  2\sigma_{od} + \frac{c_2}{L^2} + \frac{c_3}{L^3}\ .
\end{equation}
In case of the 2-parameter fits we set $c_3=0$. While the 3-parameter 
fits are somewhat unstable, consistent 2-parameter fits are limited 
to the largest three lattices. The differences between these fits 
exhibit systematic errors, which show that larger lattices are needed 
for high precision results. The results of the 2-parameter fits are 
compiled in table~\ref{tab_estimates3}, where the differences to the 
less stable 3-parameter fits are used to estimate systematic errors, 
which are, in these cases, substantially larger than statistical 
errors of the fits.

\begin{table}
\centering
\caption{\label{tab_estimates3}{Estimates of interface tensions
(yes/no refers to capillary waves, $q=3$ from Ref.~\cite{BaBe07}).}} 
\medskip
\begin{tabular}{|c|c|c||c|c|c|}
\hline\hline
 $q$ & $2\,\sigma_{od}$ -- yes & $2\,\sigma_{od}$ -- no &
 $q$ & $2\,\sigma_{od}$ -- yes & $2\,\sigma_{od}$ -- no \\ \hline
  3  & 0.001806 (35) & 0.001602 (35) & 7 & 0.1484 (15) & 0.1478 (17) \\ \hline
  4  & 0.0224 (11)   & 0.0221 (13)   & 8 & 0.1897 (56) & 0.1891 (56) \\ \hline
  5  & 0.0632 (21)   & 0.0628 (22)   & 9 & 0.2308 (87) & 0.2302 (84) \\ \hline
  6  & 0.1054 (50)   & 0.1050 (45)   &10 & 0.2628 (40) & 0.2688 (47) \\ \hline
\hline 
\end{tabular}
\end{table}

We also include in table~\ref{tab_estimates3} results from fits
without the capillary wave contribution $\ln(L)/(2L^2)$. With the 
exception of the $q=3$ case from Ref.~\cite{BaBe07}, the difference
between the two fits is always smaller than the expected error from
other sources.

\begin{figure} \begin{center}
\epsfig{file=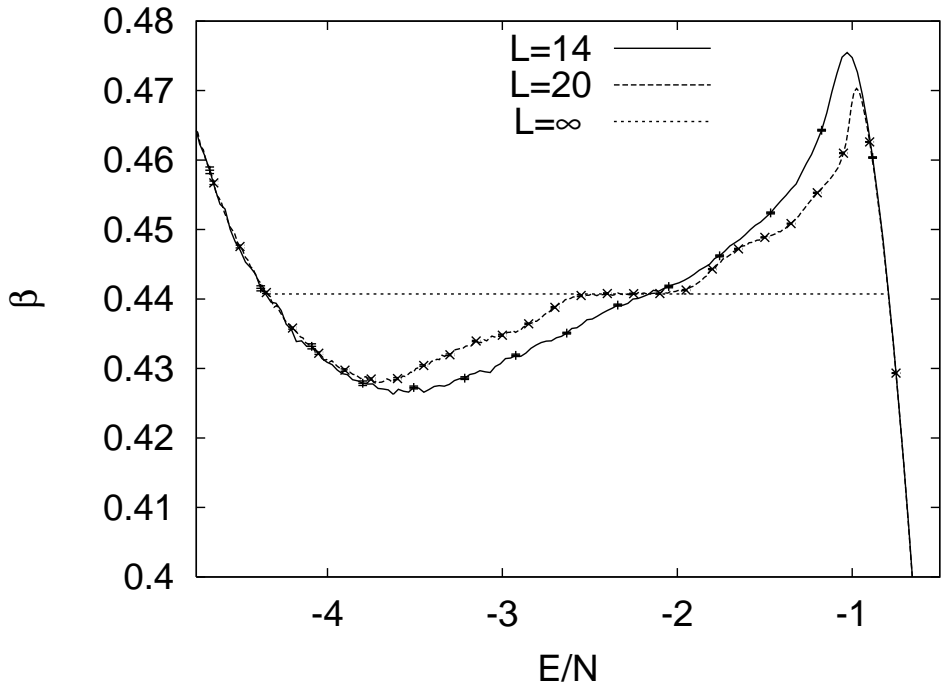,width=0.82\textwidth, angle=0}
\caption{Microcanonical temperature $b(E)$ for $q=10$.} 
\label{fig_betaE} 
\end{center} \end{figure}

The double peak histograms at first order phase transitions are
intimately related to a Maxwell construction \cite{Hu94,WJ98} for 
the inverse microcanonical temperature $b(E)$ defined by 
Eq.~(\ref{MUCAweights}). For $q=10$ and two lattice sizes this
is shown in Fig.~\ref{fig_betaE}. The areas above and below
the infinite volume line are identical and for $L=20$ one
sees that a small fraction of the curve joins this line.

The minimum $\beta^{\rm \, sp}_{\min}$ and the maximum $\beta^{\rm\, 
sp}_{\max}$ of the $\beta(E)$ curve are the inverse spinodal 
temperatures. Equilibration at $\beta$ with dissipative model~A 
(Glauber) dynamics \cite{Gl63} encounters metastability in the range 
$\beta_t<\beta<\beta^{\rm\,sp}_{\max}$ after a disordered start, 
whereas after an ordered start it encounters metastability in the 
range $\beta_t>\beta>\beta^{\rm\,sp}_{\min}$. For the $\beta(H)$ 
of a magnetic field driven phase transition \cite{BHN93} this would 
already be the entire metastability picture.
In case of the temperature driven phase transitions of Potts model it
is more complicated, because metastability after a disordered start 
persists for equilibration at $\beta>\beta^{\rm \, sp}_{\max}$ due 
to order-order domain walls, which are for $q=3$ investigated in 
Ref.~\cite{BMV04}.

In the past there may have been some hesitation in identifying 
$\beta^{\rm \, sp}_{\min}$ and $\beta^{\rm \, sp}_{\max}$, as defined 
here, with the spinodal endpoints. The reason is that their values 
agree in the infinite volume limit with $\beta_t$ \cite{Hu94,WJ98} 
as is illustrated by the dotted line in Fig.~\ref{fig_betaE}. So 
the metastability disappears in the infinite volume limit, whereas
the opposite is the case for the mean field spinodal, which is
introduced in many textbooks \cite{LB00}. However, the recent 
finite volume analysis \cite{BD08} of 
Kolmogorov-Johnson-Mehl-Avrami (KJMA) theory demonstrates that the 
mean field approach is a conceptually wrong starting point for 
describing the infinite volume limit of phase conversion. Within 
the KJMA framework one gets for $V\to\infty$ always spinodal 
decomposition \cite{spinodal} and never metastability. Our 
definitions of $\beta^{\rm\,sp}_{\min}$ and $\beta^{\rm\,sp}_{\max}$
are consistent with this picture as well as with studies of magnetic 
field driven phase transitions by Rikvold et al.~\cite{Ri94}.

\begin{figure} \begin{center}
\epsfig{file=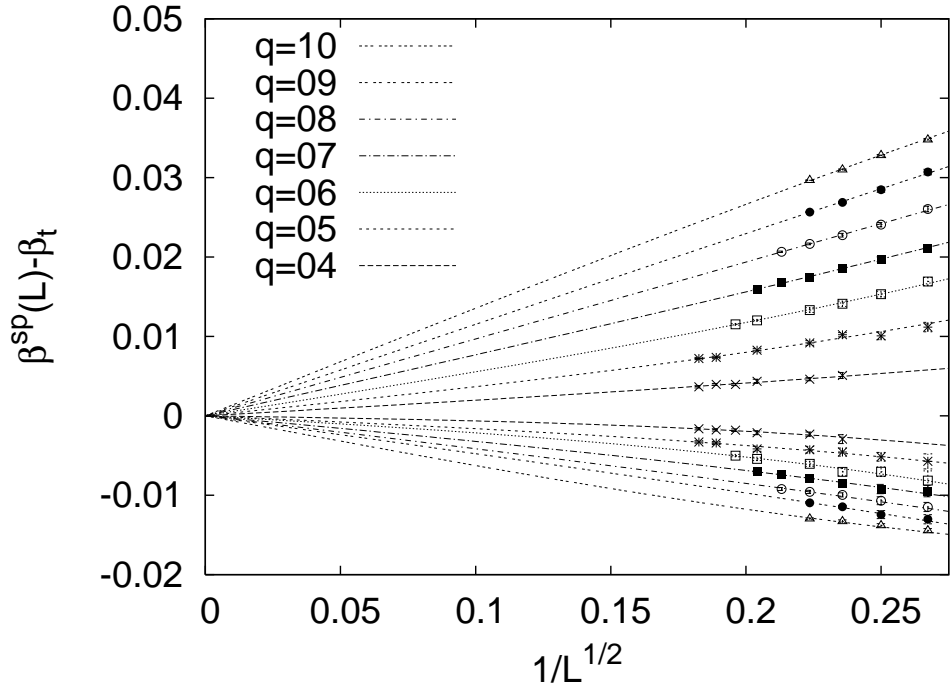,width=0.82\textwidth, angle=0}
\caption{FS behavior of the spinodals.} 
\label{fig_spinodal} 
\end{center} \end{figure}

For large $L$ the areas in the Maxwell construction are known
\cite{WJ98} to shrink $\sim 1/L$. Therefore, the leading order
2-parameter fit for $\beta^{\rm\, sp}(L)$ is
\begin{equation}\label{spinodal_fit}
  \beta^{\rm \, sp}(L) - \beta_t = \frac{a_1}{\sqrt{L}}\,
                         \left(1 + \frac{a_2}{L}\right)\,.
\end{equation}
Using $\beta_t$ from table~\ref{tab_estimates1} we show in
Fig.~\ref{fig_spinodal} the fits to this form. Together with their 
goodness of fit $Q$~\cite{BBook} the fit parameters are collected 
in table~\ref{tab_estimates4}. The $Q$ values are a bit on the high 
side, as a relatively flat $\beta(E)$ curve tends to give rather large
statistical errors for the spinodal estimates $\beta^{\rm\,sp}(L)$. In
table~\ref{tab_estimates4} this is reflected by $a_2$ parameters,
which are mainly statistical noise about zero. For the $q=3$ data 
\cite{BaBe07}, $\beta(E)$ is altogether too flat to allow for reasonably 
accurate $\beta^{\rm\,sp}(L)$ estimates (larger lattices would be needed).

\begin{table} \centering
\caption{\label{tab_estimates4}Estimates of the fit parameters of 
Eq.~(\ref{spinodal_fit}).}
\begin{tabular}{|c|c|c|c|c|c|c|}
\hline\hline
 $q$ & $a_{1,\max}$ & $a_{2,\max}$ & $Q$ & $a_{1,\min}$ & $a_{2,\min}$& $Q$ 
\\ \hline
 $4$ & 0.0193 (21)  & 1.6 (3.0) & 0.57 & -0.0053 (24)& 21 (21) & 
     0.83  \\ \hline
 $5$ & 0.0357 (20)  & 3.0 (1.5) & 0.19 & -0.0154 (19)& 5.5 (3.7)& 
     0.84  \\ \hline
 $6$ & 0.0543 (20)  & 2.07 (89) & 0.84 & -0.0200 (29)& 7.5 (4.3)& 
     0.81  \\ \hline
 $7$ & 0.0764 (22)  & 0.53 (57) & 0.67 & -0.0315 (27)& 2.3 (2.0)& 
     0.95  \\ \hline
 $8$ & 0.0969 (10)  &-0.01 (19) & 0.85 & -0.0412 (27)& 0.8 (1.4)& 
     0.69  \\ \hline
 $9$ & 0.1154 (19)  &-0.14 (30) & 0.58 & -0.0475 (25)& 0.6 (1.1)& 
     0.82  \\ \hline
 $10$ & 0.13631 (93)&-0.58 (12) & 0.05 & -0.0639 (19)& -2.00 (41)  & 
     0.21  \\ \hline
\hline 
\end{tabular} \end{table}

\section{Summary and Conclusions} \label{sec_sum}

For 3D, $q$-state Potts models in the range $q=4,\dots,10$ we have 
estimated a number of observables by multicanonical MCMC calculations 
and supplemented them with $q=2$ and $q=3$ results from the literature. 
Transition temperatures, latent heats and energy endpoints of the 
phases are given in table~\ref{tab_estimates1}, entropy and free 
energy values in table~\ref{tab_estimates2}. 

Less accurate are our interface tension estimates of 
table~\ref{tab_estimates3}. They could possibly be improved
by using simulation techniques similar to those, which led to
high-precision estimates of the order-order interface 
tension in the 3D Ising model~\cite{CHP07}.

Minima and maxima of the microcanonical inverse temperature curve
$b(E)$ are identified as adequate definition of spinodal endpoints.
As expected \cite{BD08,Ri94} the thus defined regions of metastability
disappear in the infinite volume limit.

We hope that future investigations of 3D first order phase transitions
will benefit from the results collected in this paper.
\bigskip

\noindent {\bf Acknowledgements:} This work was in part supported by
DOE grants DE-FG02-97ER-41022 and DE-FC02-06ER-41439 and by NSF grant
0555397.


\begin{thebibliography}{99}

\bibitem{Po52} R.B. Potts, Proc. Cambridge Philos. Soc. 48 (1952) 106.

\bibitem{BBook} B.A. Berg, {\it Markov Chain Monte Carlo Simulations
                and Their Statistical Analysis}, World Scientific, 
                Singapore, 2004.

\bibitem{Wu82} F.Y. Wu, Rev. Mod. Phys. 54 (1982) 235.

\bibitem{Ba73} R.J. Baxter, J. Phys. C 6 (1973) L445.

\bibitem{BoJa92} C. Borgs and W. Janke, J. Phys. I France 2 (1992) 
                 2011 and references therein.

\bibitem{BeNe92} B.A. Berg and T. Neuhaus, Phys. Rev. Lett. 68 (1992) 9.

\bibitem{EdAn75} S.F. Edwards and P.W. Anderson, J. Phys. F 5 (1975)
                 965.

\bibitem{Bi97} K. Binder, {\it Quadrupolar Glasses and Random Fields},
in {\it Spin Glasses and Random Fields}, A.P. Young (editor), World
Scientific, Singapore 1997.

\bibitem{Vi77} J. Villain, J. Phys. C 10 (1977) 1717.

\bibitem{SY82} B. Svetitsky and L.G. Yaffe, Nucl. Phys. B 210
               (1982) 443.

\bibitem{GKB89} R.V. Gavai, F. Karsch, and B. Petersson, Nucl. Phys. B
                322 (1989) 738.

\bibitem{FMOU90} M. Fukugita, H. Mino, M. Okawa, and A. Ukawa,
                 J. Stat. Phys. 59 (1990) 1397.

\bibitem{ABV91} N. Alves, B.A. Berg and R. Villanova, Phys. Rev. B
                43 (1991) 5846.

\bibitem{Schm94} M. Schmidt, Z. Phys. B 95 (1994) 327.

\bibitem{JV97} W. Janke and R. Villanova, Nucl. Phys. B 489 (1997) 679.

\bibitem{KaSt00} F. Karsch and S. Stickan, % {\it The three-dimensional,
                 Phys. Lett. B 488 (2000) 319.
% 3D 3-state Potts model in and external field}, --325; hep-lat/7019.

\bibitem{Fa07} R. Falcone, R. Fiore, M. Gravina, and A. Papa, Nucl.
               Phys. B 767 (2007) 385. % hep-lat/0612016.

\bibitem{BaBe07} A. Bazavov and B.A. Berg, Phys. Rev. D 75 (2007) 094506.

\bibitem{GN02} A. Gendia and T. Nishino, Phys. Rev. E 65 (2002) 046702.
               % cond-mat/0102425.

\bibitem{Ha05} A.K. Hartmann, Phys. Rev. Lett. 94 (2005) 050601. 
               % cond-mat/0410583.

\bibitem{HJ06} M. Hellmund and W. Janke, Phys. Rev. E 74 (2006) 051113.
               % cond-mat/0607423.

\bibitem{HS95} B. Hesselbo and R. Stinchcombe, Phys. Rev. Lett. 74 
               (1995) 2151.

\bibitem{THT04} S. Trebst, D.A. Huse, and M. Troyer, Phys. Rev. E 70
                (2004) 046701.

\bibitem{NH03} T. Neuhaus and J.S. Hager, J. Stat. Phys. 113 (2003) 47.

\bibitem{WL01} F. Wang and D.P. Landau, Phys. Rev. Lett. 86 (2001) 2050.

\bibitem{BHN93} B.A. Berg, U.H. Hansmann, and T. Neuhaus, Z. Phys.
                90 (1993) 229.

\bibitem{TB96} A.L. Talapov and H.W.J. Bl\"ote, J. Phys. A: Math.
               Gen 29 (1996) 5727.

\bibitem{CLB86} M.S.S. Challa, D.P. Landau, and K. Binder, Phys.
                Rev. B 34 (1986) 1841.

\bibitem{Bi82}  K. Binder, Phys. Rev. A 25 (1982) 1699.

\bibitem{BZ} E. Br\'ezin and J. Zinn-Justin, Nucl. Phys. B 
             257 (1985) 867.

\bibitem{GF} M.P. Gelfand and M.E. Fisher, Physica A 166 (1990) 1.

\bibitem{Mo} J.J. Morris, J. Stat. Phys. 69 (1991) 539.

\bibitem{BNB94} A. Billoire, T. Neuhaus and B.A. Berg, Nucl. Phys. B 
                413 (1994) 795.

\bibitem{Hu94} A. H\"uller, Z. Phys. B 95 (1994) 63.

\bibitem{WJ98} W. Janke, Nucl. Phys. B (Proc. Suppl.) 63A-C (1998) 631.

\bibitem{Gl63} R.J. Glauber, J. Math. Phys. 4 (1963) 294. Model~A in
               the classification of P.M. Chaikin and T.C. Lubensky,
               {\it Principles of condensed matter physics}, Cambridge
               University Press, Cambridge 1997, Table~8.61.1, p.467.

\bibitem{BMV04} B.A. Berg, H. Meyer-Ortmann, and A. Velysky, Phys.
                Rev. D 70 (2004) 054505; A. Bazavov, B.A. Berg, and
                A. Velytsky, Phys. Rev. D 74 (2006) 014501.

\bibitem{LB00} E.g., D.P. Landau and K. Binder, {\it A Guide to Monte 
               Carlo Simulations in Statistical Physics}, Cambridge 
               University Press 2000, p.41.

\bibitem{BD08} B.A. Berg and S. Dubey, Phys. Rev.  Lett. 100 (2008)
               165792.

\bibitem{spinodal} Here we use the terminology ``spinodal decomposition''
               in a broader sense than some statistical physicists do.

\bibitem{Ri94} P.A. Rikvold, H. Tomita, S. Miyashita, and S.W. Sides,
               Phys. Rev. E 49 (1994) 5080; M.A. Novotny, G. Brown,
               and P.A. Rikvold, J. Appl. Phys. 91 (2002) 6908.

\bibitem{CHP07} M. Caselle, M. Hasenbusch, and M. Panero, JHEP 9
                (2007) 117 and references given therein.
                
\end{thebibliography}
\end{document}